# Synthetic Hachimoji DNA Sequencing with Graphene Nanodevice

Rameshwar L. Kumawat* and Biswarup Pathak*

Department of Chemistry, Indian Institute of Technology (IIT) Indore, Indore, Madhya Pradesh, 453552, India

*E-mail: rameshwarlal1122@gmail.com; biswarup@iiti.ac.in

**ABSTRACT:** Based on combined density functional theory and non-equilibrium Green's function quantum transport studies, we have demonstrated quantum interference (QI) effects on the transverse conductance of Hachimoji (synthetic) nucleic acids placed between the oxygen-terminated zigzag graphene nanoribbon (O-ZGNR) nanoelectrodes. We theorize that the QI effect could be well preserved in π-π coupling between a nucleobase molecule and the carbon-based nanoelectrode. Our study indicates that QI effects such as anti-resonance or Fano-resonance that affect the variation of transverse conductance depending on the nucleobase conformation. Further, a variation of up to 2-5 orders of magnitude is observed in the conductance upon rotation for all the nucleobases. The current-voltage (I-V) characteristics results suggest a distinct variation in the electronic tunnelling current across the proposed nanogap device for all the five nucleobases with the applied bias voltage. The different rotation angles keep the distinct feature of the nucleobases in both transverse conductance and tunnelling current features. Both features could be utilized in an accurate synthetic DNA sequencing device.

**INTRODUCTION**

In biology, ribonucleic acid (RNA), deoxyribonucleic acid (DNA), and proteins are large linear biopolymer that contain the biological and genetic data of all living organisms at the molecular level.[1,2] These linear biopolymers comprise of monomeric units such as DNA nucleotides of one of the four naturally occurring nucleobases connected via a sugar group and a phosphate backbone. The sequence of all four nucleotides serves as template material for protein creation and synthesis and the transmission of genetic and biological information in our body.[3,4] Very recently, Malyshev and co-workers have demonstrated a novel way in artificial (or synthetic) biology by presenting exogenous unnatural nucleobase pairs into a living organism's DNA.[5] They have shown the viability of promulgating an augmented genetic alphabet. After that, the field has demonstrated the creation of a semi-synthetic DNA and RNA nucleobases system based explicitly on an eight-letter building block known as Hachimoji.[6] Along with natural DNA nucleobase hydrogen (H)-bond pairs, Hachimoji DNA also forms two additional H-bond patterns resulting from five new synthetic nucleobases, namely: B, P, rS, S, and Z as shown in **Figure 1(a)**. These are the synthetic (unnatural) nucleobases synthesized by Hoshika and co-workers.[6] In likeness to the natural double-stranded (ds)-DNA, the artificial DNA nucleobase S bonds with B, and P bonds with Z, while B bonds with rS in the unnatural RNA system. These DNA building blocks are also P and B, which are analogues of *purine* system, and S and Z, which are analogues of *pyrimidine* system. These duplexes form B:S and P:Z pairs in unnatural DNA. Like natural DNA, Hachimoji DNA, though unnatural (artificial), can support the evolution of organisms. Also, Hachimoji DNA can be used to develop and advance clean diagnostics for killer human disease, self-assembling nanostructures, in retrievable DNA digital information storage, DNA barcoding, and to make proteins with novel drugs and additional amino acids.[6]

The experimental proof of artificial genetic characters presents a challenge to low-cost, label- and amplification-free, rapid, and controlled next-generation sequencing (NGS) techniques such as nanogap, nanopore, and nanochannel-based smart sensors for healthcare applications.[3,4,7–15] In NGS, DNA nucleobases can electrophoretically be passed through nanogaps or nanopores, which changes ionic or tunneling/transverse currents which may be used in real-time sequencing. On the other hand, nanochannel-based devices are also quite promising for NGS applications. This could hold each nucleobase firmly while the target nucleobase is being sequenced/identified. Although biological nanopores-based devices have several advances and have already been commercialized for DNA sequencing, some disadvantages still need to be resolved, for example, high noise in experimental measurements, uncontrolled translocation, and high cost, among others.[16–18] Therefore, to realize a low-cost, label- and amplification-free, rapid, and controlled NGS, solid-state materials-based nanopores, nanogaps, and nanochannels are found to be promising. In this perspective, nanopores in solid-state materials have been found quite promising for next-generation DNA sequencing devices. Compared to other solid-state materials, atomically thick solid-state graphene has attracted significant attention due to its excellent mechanical, structural, electronic, and quantum transport properties.[4,11–13,19] In addition, graphene thickness is narrower than the distance between two nucleotides in a DNA strand which allows for a single-nucleobase resolution at the molecular level. Consequently, to assess the applicability of a graphene-based nanopore device to sequence DNA, experimental and theoretical studies have examined the effect of all four nucleotides inside a nanometer-size nanopore or nanogap-based device on ionic as well as tunnelling/transverse current signals.

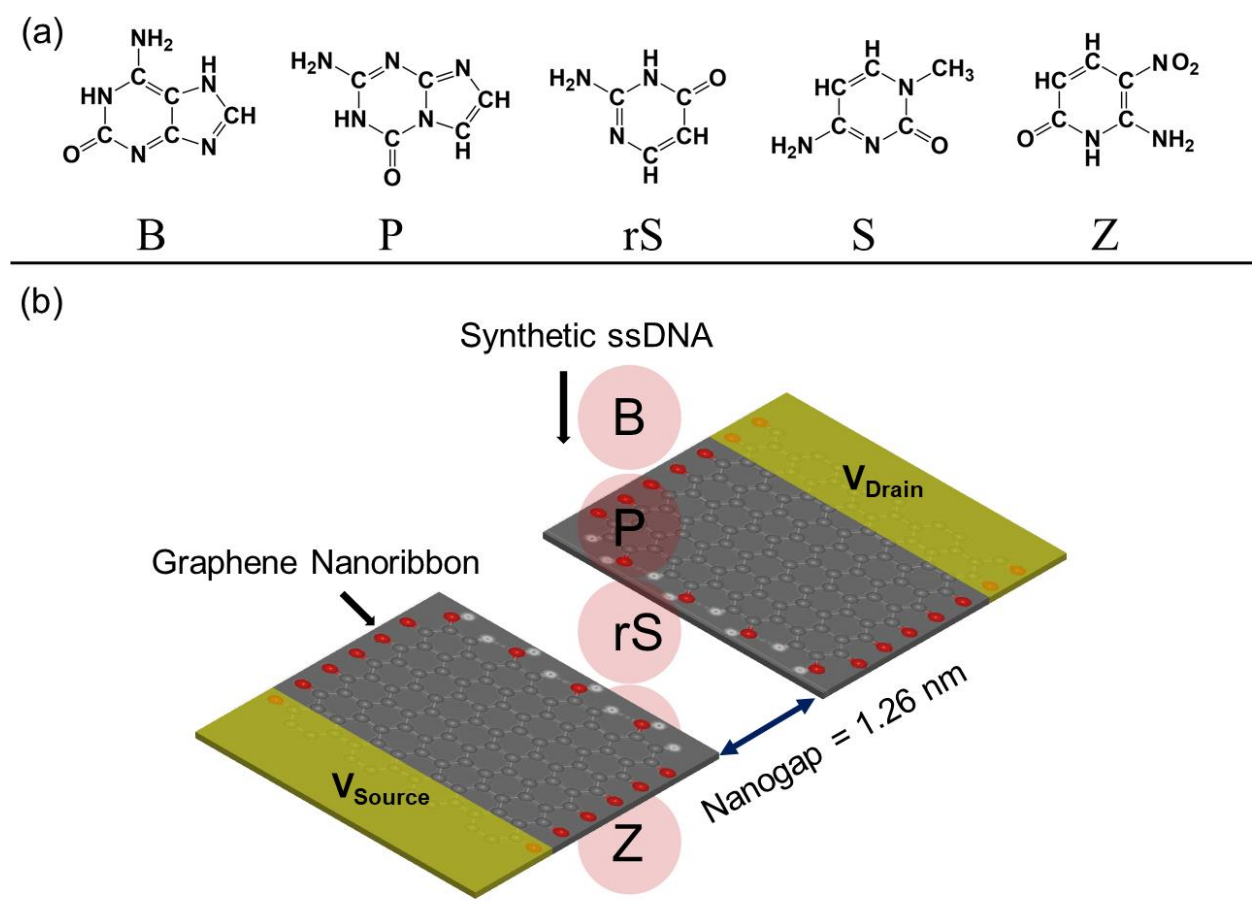

**Figure 1.** (a) Chemical structures of Hachimoji DNA nucleobases (B, P, rS, S, and Z). (b) Schematic simulation setup of a proposed graphene nanogap device with target nucleobase molecule.

Graphene nanoribbon nanoelectrodes have been proposed theoretically and experimentally for various applications in the past few years. The edge-chemistry modified zigzag graphene nanoribbons (ZGNR) include CN, CF, CH, $CH_2$ (H passivation's), COH, CO, and $C_2O$ (oxidations).[19–24] In a report, Lee and co-workers have shown the band structure of H or $H_2$ edge-hydrogen terminated ZGNRs as well as CO, COH, or $C_2O$ edge-oxidized ZGNRs.[25] They have reported that the OH group yields nearly the same band structure as the $sp^2$ hybridization of H-edge. The ketone (CO) and ether ($C_2O$) groups result in band structure analogous to those of $sp^3$ hybridization of $H_2$-edge. The O-edge gives an energetically more favorable system than the H-edge with the metallic nature of oxidized GNRs. Motivated by all these reports, we investigate the electronic structure and quantum transport properties of the oxygen-terminated ZGNR (O-ZGNR) nanogap setup. The nanogap edges are terminated by alternating OH and H groups [**Figure 1(b)** and **S1**]. This is done to preserve the delocalized electronic wavefunction or π-conjugation of the O-ZGNR setup. We believe that this hypothetical device could be a more feasible and realistic nanogap device compared to the hydrogen (H) and nitrogen (N)-functionalized device when we consider the wet-fabrication process. Further, we consider this hypothetical setup to check the applicability for NGS of Hachimoji DNA. For this, we have considered all five Hachimoji DNA nucleobases (nucleobase: B, P, rS, S, and Z) inside the proposed nanogap setup. The first-principles density-functional-theory (DFT) approach is used to study the electronic structure, interaction energy ($E_i$), the density of states (DOS), charge density difference (CDD) plots of O-ZGNR (with and without nucleobases) setup. The first-principles non-equilibrium Green's function (NEGF) method is used to study the quantum transport properties of Hachimoji DNA nucleobases while placed between two closely spaced O-ZGNR nanoelectrodes. The transverse transmission ($G_0$) function and tunneling current-voltage (I-V) characteristics may be utilized for the individual identification of all five nucleobases. The $G_0$ spectra and I-V characteristics at different rotation angles have also been studied. In addition, we look at quantum coherence features of electronic transport devices, i.e., *destructive quantum interference* (DQI) between two electronic paths.

**RESULTS AND DISCUSSION**

We commence the discussion of results with the proposed O-ZGNR device characteristics. With this, first, we refer to the structural and electronic features of the O-ZGNR+nucleobase (nucleobase: B, P, rS, S, and Z) systems, in which one of the five Hachimoji DNA nucleobases is being placed. Second, we refer to the electronic transverse $G_0$ function and tunneling I-V characteristics features across the O-ZGNR+nucleobase device. The electronic transverse $G_0$ function, tunnelling current, and pathways to improve it are vital for an error-free sequencing of Hachimoji DNA nucleobases.

Investigation of all fully relaxed configurations discloses that nucleobases (B, P, S, and Z) remain in-plane with the O-ZGNR device, while rS nucleobase assumes a tilted configuration (**Figure S1(a-e)**). To evaluate the energetic stability of nucleobases in the O-ZGNR device, the $E_i$ values are computed using equation 1. The $E_i$ values may help as an identifier of Hachimoji DNA nucleobases, as it can be directly related to their translocation time ($\tau$). The evaluated $E_i$ values are found to be -1.71, -0.76, -1.01, -0.86, and -1.70 eV for B, P, rS, S, and Z nucleobases, respectively. They are following the sequence: B ≥ Z > rS > S > P. These values are in the range from -1.71 to -0.76 eV, nearly comparable to those of the natural DNA nucleobases in other boron-carbide (BC$_3$) and graphene-based nanogap systems.[26–29] In addition, these $E_i$ values specify that the existence of Hachimoji DNA nucleobases (synthetic DNA) in modified samples may not affect the capability of the graphene device and other nanogap devices to sequence them adequately. It is found that the synthetic DNA nucleobases (B, P, rS, and S) are stabilized at nearly similar distances from edges of nanogap with a deviation of ~0.1 Å for B, rS, and P nucleobases and ~0.2 in the case of S and rS nucleobases. Although in the case of Z, it is stabilized at a higher distance from the nanogap edges with a deviation of ~1.2 Å compared to other nucleobases. This could be because of the smaller size of the Z molecule. For nearly similar distances, the $E_i$ values could be determined by the weak H bonds formed between molecule and nanoelectrode edge. The weak formation of one O-H⋯O bond is found for all five nucleobases. However, one O-H⋯O and one O-H⋯N bond for the B nucleobase is observed. Likewise, it is observed that 3O atoms could form weak bonding with 3H atoms of nanoelectrode for the Z nucleobase. In addition, the interaction between O and H-bonds could be more consistent than that of N and H-bonds, which again depends on several parameters. One may be due to the higher electronegativity of the O (~3.5) atom than the N (~3.0) atom. This could be why B and Z nucleobases show relatively high $E_i$ values than the other three nucleobases (P, rS, and S). Therefore, these results show that the $E_i$ values are dominated by charge-transfer interactions. Nevertheless, the formation of a weak H-bond depends on the molecular orientations. Thus, our analysis suggested that the formation of a temporarily weak H-bond between molecule and nanoelectrode edge can stop each target nucleobase for a short period when it translocates over the O-ZGNR nanogap device. It results in enhanced coupling between the molecule and nanoelectrode and may also stabilize the translocating DNA against the orientation fluctuation, which can lower the noise ratio during experimental measurements of transverse conductance and tunneling current. Further, the CDD plot analysis shows the charge-transfer interaction between the target molecule and nanoelectrode edges. The charge fluctuates around the nanoelectrode edges adjacent to the nucleobases, as shown in **Figure 2**(upper panel). So, such charge redistribution facilitates the way for charge transport in the proposed O-ZGNR device through target nucleobase.

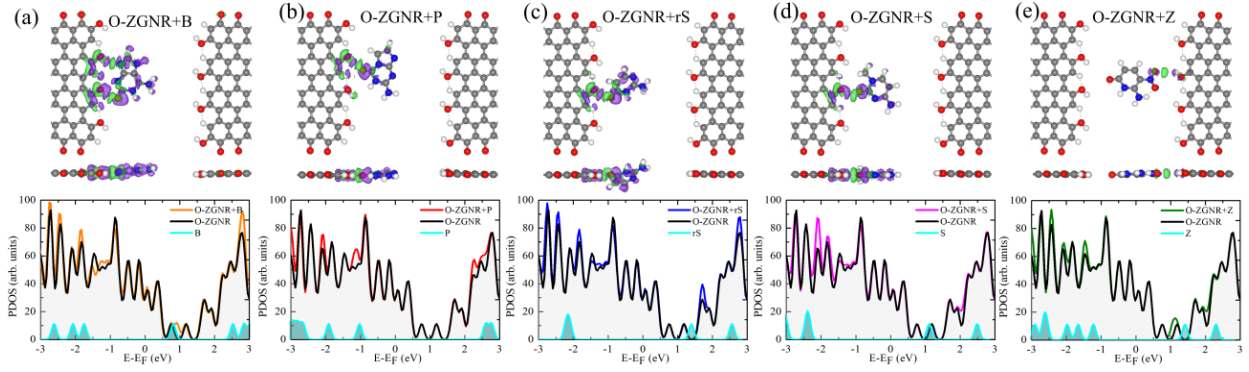

**Figure 2.** (**a-e**) Top and side views of the charge density difference maps (iso-surface value: 0.005 e/Å$^3$) for O-ZGNR+nucleobase (nucleobase: B, P, rS, S, and Z) systems (upper panel) and electronic DOS of the O-ZGNR and O-ZGNR+nucleobases (nucleobases: B, P, rS, S, and Z) systems and the PDOS for each target nucleobase (down panel). The violet iso-surface signifies an electron gain and the green iso-surface signifies an electron loss. The Fermi-level is fixed to zero.

Further, the effect of DNA molecule interaction on the electronic structure of the O-ZGNR is investigated. The computed electronic total/projected DOS (T/PDOS) of the O-ZGNR and O-ZGNR+nucleobase, together with the PDOS of all five nucleobases, are shown in **Figure 2**(down panel). The weak interaction between molecule and nanoelectrode leads to changes in the electronic DOS of the O-ZGNR. It can be seen that in the given energy region $\pm 3$ eV, the target nucleobases placed inside the two closely spaced nanoelectrodes give several molecular states around the Fermi-level of the O-ZGNR device. However, the interaction of Hachimoji nucleobases (B, P, rS, S, and Z) does not affect the metallic nature of the device while placed inside the O-ZGNR nanoelectrodes. Furthermore, it is found that B and Z nucleobases have the maximum number of available molecular states in the $\pm 3$ eV energy region, while the other nucleobases have the least number of molecular states in the same energy region. However, each nucleobase has several available molecular states, suggesting that each target nucleobase molecule is coupling well with the O-ZGNR nanoelectrodes. This reveals the optimum coupling between molecule and nanoelectrodes, which is neither strong nor weak. On the other side, a more robust coupling is desired to enhance the transverse $G_0$ function and tunnelling current signals, which can be used to detect each nucleobase while translocating through the O-ZGNR nanogap device.

Next, to understand Hachimoji DNA sequencing performance of the O-ZGNR device, the transverse $G_0$ spectra and tunneling I-V characteristics by using a two-probe device (**Figure 1(b)**) are investigated, as discussed in the method section. The computations are performed for the O-ZGNR device (with and without nucleobases). The resulting $G_0$ spectra as a function of energy (E) at zero-bias for an energy window of E-$E_F$ = $\pm 3.5$ eV is shown in **Figure 3(a)**. The $G_0$ spectra curve of an open nanogap device is labelled as "O-ZGNR", which serves as the reference to identify the prospect to detect a specific nucleobase. It is found that all $G_0$ spectra curves corresponding to a specific nucleobase in the nanogap device exhibit a reduction in the $G_0$ function compared to the reference bare "O-ZGNR" device (solid black line) for energy values below/above the Fermi-level. The reduction in the $G_0$ function could be assigned to the electrostatic interaction nature of the O-ZGNR device with the negatively charged sites of Hachimoji nucleobase (i.e., O site). These negatively charged sites are in the locality of the graphene-nanoelectrode, which establishes a negative electrostatic potential and ushers to the aforesaid statement. The same trend has been observed for the graphene-based nanopore device when natural DNA nucleobases are located inside the nanopore.[30] Further, we find that at an energy E-$E_F$ = 0 eV (**Figure 3(a)**), the $G_0$ i.e., conductance magnitudes follow a sequence: O-ZGNR > Z > S > rS $\cong$ B $\cong$ P. This suggests that Z and S nucleobases have better conductance (Z > S) values than the other three (rS $\cong$ B $\cong$ P) nucleobases at the Fermi-level.

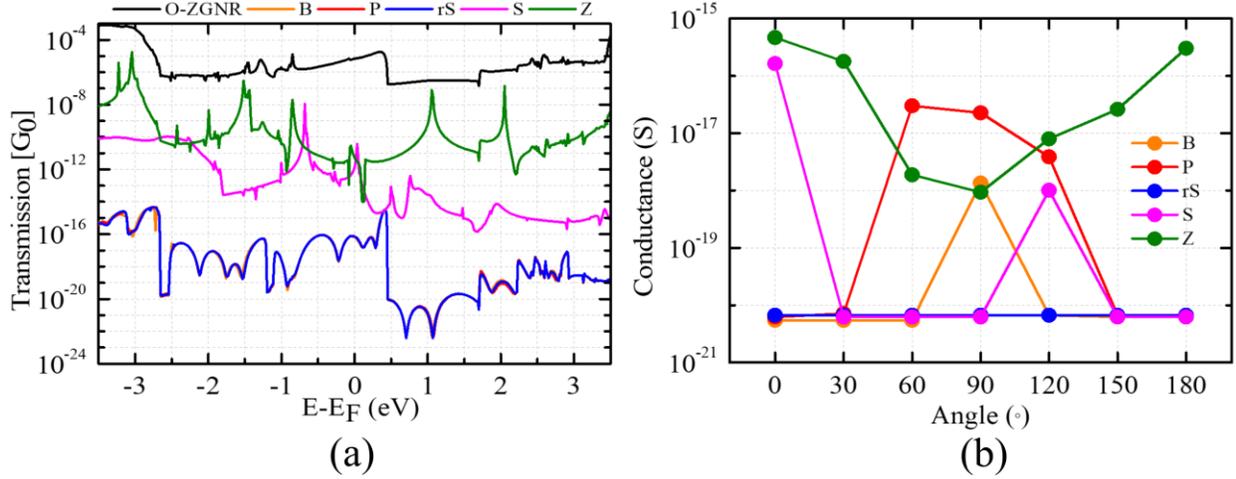

**Figure 3.** (a) Transmission ($G_0$; at zero-bias) spectra curve plotted on a logarithmic scale for the empty O-ZGNR nanogap device and the O-ZGNR+nucleobase (nucleobases: B, P, rS, S, and Z) systems. Here the $G_0$ curve for the reference nanogap device is shown by the black-solid-line, signifying that the $G_0$ spectra features of Hachimoji DNA nucleobases resemble vacuum tunnelling. (b) Transverse tunnelling conductance of B, P, rS, S, and Z nucleobases while located inside the nanogap device, at different rotations ranging from 0° to 180°. The conductance values are obtained from the $G_0$ curve at the Fermi-level and then multiplied by 77.6 µS.

Further, we have investigated the dependency of the $G_0$ function on the rotations of isolated Hachimoji DNA nucleobases. For this, the rotation of all five nucleobases is performed inside the nanogap device. Considering the sufficient nanogap size and the possible conformation of nucleobases, we rotated each nucleobase from 0° to 180° in steps of 30° around the *x*-axis in the *yz*-plane [**Figure S2-S6**]. All five nucleobases are considered to be freely rotating within this range of angles. In **Figure 3(b)**, we show the conductance of all five nucleobases at different rotation angles of 0° to 180°, as calculated from the $G_0$ at the Fermi-level. A variation of up to 2-5 orders of magnitude is observed in the conductance upon rotation. A similar interpretation has also been reported for the graphene-based device with natural DNA.[30] It is observed that the magnitude and variation of the Z nucleobase conductance are significantly larger compared to other nucleobases. Z nucleobase ranges from 0.94 aS (atto Siemens) at 90° to 466.90 aS at 0°. It is noted that the order of the conductance for all five nucleobases changed with respect to nucleobases rotation angles, and we find that Z nucleobase could be robustly identified from the other four nucleobases. Nevertheless, if we fix the nucleobase angle, it may be possible to detect some of these nucleobases to some extent from the conductance orders, for example, Z >> P > B ≥ S > rS for 90° to 180° orientation. These results show that all five nucleobases in principle possible to detect while located inside the O-ZGNR nanogap device. Herein, it is essential to note that conductance order is not the influence of interaction (binding) strength but rather the different Hachimoji nucleobases that control the conductance order. In addition, nanoelectrode edges + molecule coupling (molecular orbitals) plays a significant role, which we have discussed in the forthcoming paragraph.

Next, we studied the $G_0$ spectra to understand the effect of conformational fluctuations for all five nucleobases. In **Figure 4(a-e)**, we show the logarithmic plot of $G_0$ spectra as a function of energy for $E-E_F = \pm 3.5$ eV. It is observed that the empty nanogap device has integer-like $G_0$ values around the Fermi-level corresponding to the number of available conduction channels. The $G_0$ spectra give sharp resonance peaks/dips (symmetric/asymmetric) in the presence of target nucleobase. This could be due to molecular orbital overlap between nanoelectrode and molecule. These resonance peaks/dips can be controlled by the application of an external gate voltage. Furthermore, it has been noted that the rotation of the target molecule inside the nanogap with different angles leads to change in the $G_0$ spectra of O-ZGNR+nucleobase (nucleobase: B, P, rS, S, and Z). The peak position, width, and height (shifts upward/downward) with respect to Fermi-level (Figure 5). This happens due to the change in the coupling between nanoelectrode edges+nucleobases. The difference in $G_0$ spectra magnitude has also been observed in all cases except for O-ZGNR+rS system.

Further, to understand the changes in the $G_0$ spectra and resonance peaks, we have plotted the highest occupied molecular orbitals (HOMOs) and lowest unoccupied molecular orbitals (LUMOs) at the energy level of the most characteristics sharp $G_0$ resonance peaks. In each case, a strongly mixed molecular orbital (MO) between the nanoelectrode and nucleobase is found whose nucleobase part well corresponds to the frontier molecular orbitals (FMOs) of the pristine nucleobase molecule, as shown in **Figure 5**. For the O-ZGNR+B system (at 0°, 30°, 60°, 150°, and 180°), two *anti-resonance* or *Fano-resonance* peaks appear between 0.5-1.1 eV above the E-$E_F$, as shown in **Figure 4(a)**. Several other reports have also been demonstrated such anti-resonance peaks in the natural DNA and aromatic hydrocarbon molecules and attributed them to *destructive quantum interference* (DQI) between two electronic paths. We have computed the electronic projected DOS to understand such features in our proposed device. The projected DOS results (**Figure S7**) show that $C_{px+py}$ and $O_p$ orbitals of nanoelectrode + molecule are interacting, and the resulting π-π interactions could be the reason for such anti-resonance features. Therefore, our results confirm that the *quantum coherence* is preserved in π-π interactions between the nanoelectrode and channel.[46] The same behavior can also be observed in three more systems [O-ZGNR+P (at 0°, 30°,150°, 180°); **Figure 4(b)**, O-ZGNR+rS (at 0°, 30°, 60°, 90°, 120°, 150°, 180°); **Figure 4(c)**, and O-ZGNR+S (at 30°, 60°, 180°); **Figure 4(d)**].

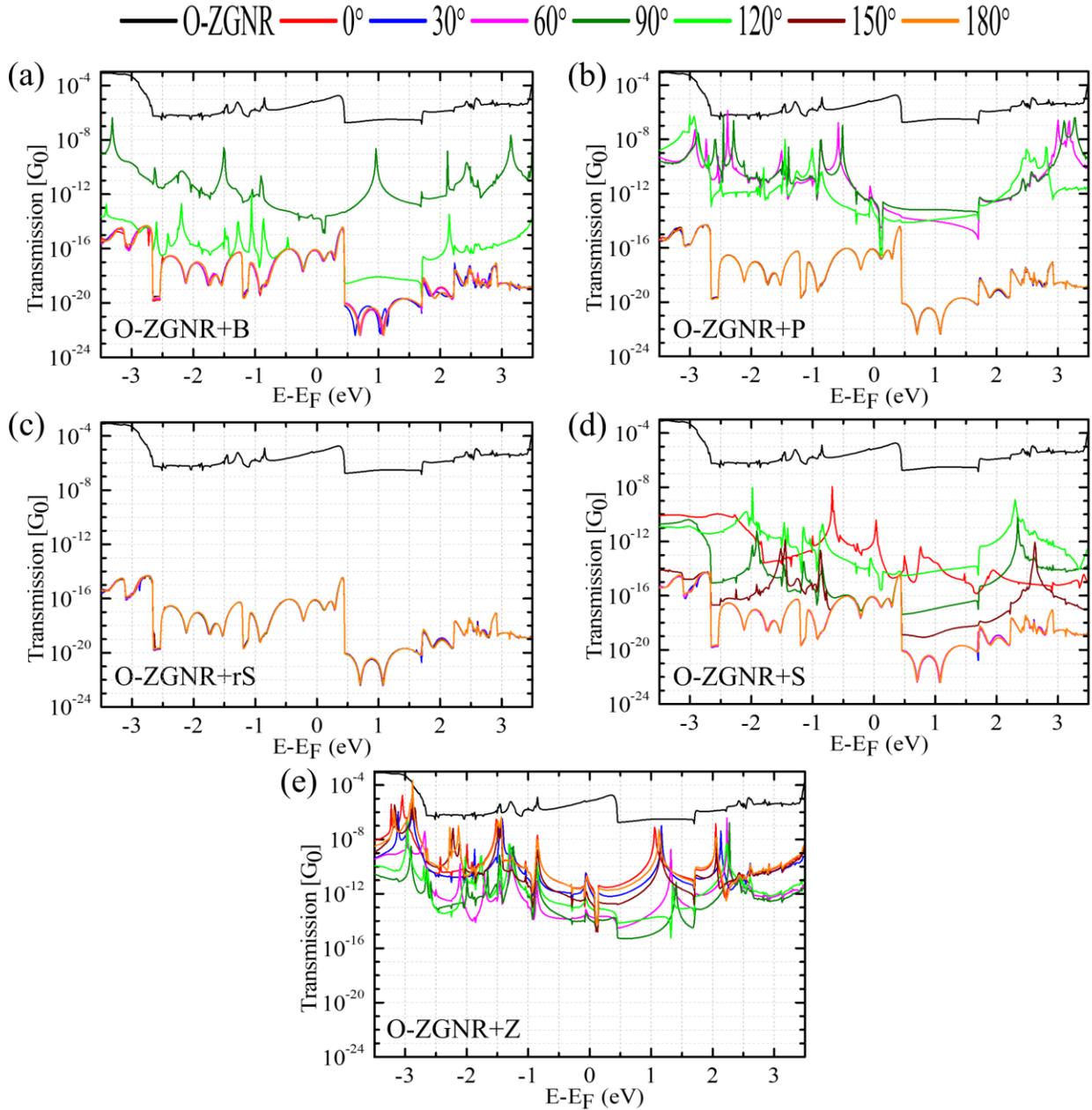

**Figure 4.** (a-e) Quantum interference effects and change in $G_0$ spectra as a function of both energy and rotation angle for all the five nucleobases. The rotation is performed in the yz-plane along the x-axis from 0° to 180° in the steps of 30° with respect to their original positions (0°).

Another important feature in the $G_0$ spectra is the appearance of a sharp resonance peak below the E-$E_F$ for the B, P, S, and Z nucleobases, as shown in **Figure 4**. By comparing the obtained $G_0$ spectra for *purine* (P and B) and *pyrimidine* (S and Z) nucleobases, we see two resonance peaks at different orientations for all four nucleobases below the E-$E_F$. For the O-ZGNR+B system, we see that the resonance peak at ~-1.50 eV (at 90° orientation) corresponds to the HOMO level of nucleobase, making a significant contribution to the $G_0$ function. From the MO picture presented in **Figure 5**, this state could be recognized to be induced-states on the O-ZGNR nanoelectrode rather than being a molecular state. Such a feature of having induced-states contributing to the $G_0$ function is a vital feature of our proposed setup. It could be contrasted to purely molecular states that have been discussed in the H-functionalized graphene nanogap by Prasongkit and co-workers.[12] Additionally, we also see the molecular state at ~0.96 eV (**Figure 4(a)** and **Figure 5**), contributing to the $G_0$ function. Likewise, in the O-ZGNR+P system, we see that the resonance peak at ~-0.58 eV (at 60° orientation, **Figure 4(b)** and **Figure 5**) is attributed to a HOMO level of the given nucleobase. On the other hand, a state at ~3.00 eV also contributes to the $G_0$ function. This state is associated with the LUMO level of the nucleobase. Furthermore, the HOMO resonance peaks can be seen at ~-0.84 eV (at 90° orientation for S and 0° for Z nucleobase; **Figure 4(d-e)** and **Figure 5**) for both nucleobases. Alike B nucleobase, we also find the induced states on the nanoelectrode for both S and Z nucleobases localizing on the right and left sides of nanoelectrode, contributing to the $G_0$ function significantly. The LUMO resonance peaks located at ~2.35 eV and ~1.06 eV for S and Z respectively are purely molecular states, as shown in **Figure 5**. Therefore, it can be noted that in all the four cases, the LUMOs corresponding to resonance peaks are found to be localized on the *purine* and *pyrimidine* nucleobases. We anticipate that these molecular states as well as induced states, contribute significantly to the I-V characteristics. Hence, from the foregoing discussion, we have concluded that molecular states of the nucleobases and the induced-states on the O-ZGNR nanoelectrode play a vital role in the $G_0$ functions.

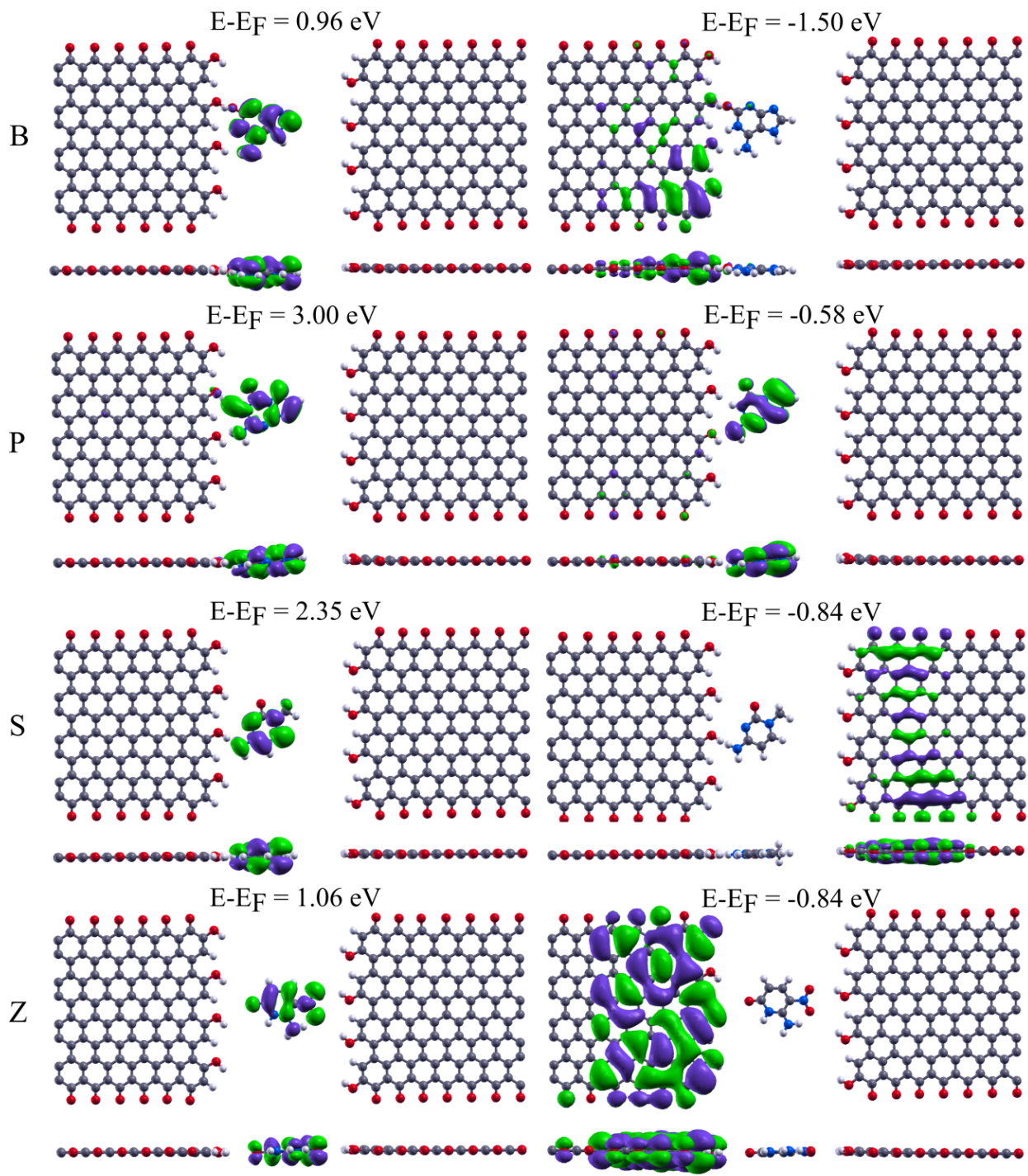

**Figure 5.** Molecular orbitals (MOs) responsible for the transmission resonance peaks (with respective energies values) are shown for nanogap+nucleobase (nucleobase: B, P, S, and Z) systems.

Next, we computed the electronic tunnelling current across the O-ZGNR nanogap device under the applied bias voltage across the nanoelectrodes in the presence of target nucleobase. The corresponding I-V characteristic curves for all the five nucleobases are shown in **Figure 6**. Herein, we focused on the nanogap's initial configuration: no rotation of the nucleobases inside the nanogap. The first inspection of **Figure 6(a)** discloses the distinct I-V pattern for S and Z nucleobases. However, the other three nucleobases (B, P, rS) have the same current values. To understand these I-V characteristics, we turn our attention to the electronic $G_0$ spectra for all five nucleobases (**Figure 3(a)**). Clearly, distinct $G_0$ patterns and resonance peaks are revealed for S and Z nucleobases, denoting that both the nucleobases can be distinguished from each other through the $G_0$ patterns and resonance peaks. In contrast, the $G_0$ patterns and resonance peaks are revealed the same for B, P, and rS nucleobases, denoting that these nucleobases cannot be distinguished from each other either through conductance nor I-V values.

We further wanted to understand how the proposed O-ZGNR nanogap device features change with the relative orientation of the nucleobases. These changes are predicted to influence the weak H- and O-bonding between the nanoelectrode edges and the nucleobase, thus the $G_0$ spectra and I-V characteristics across the O-ZGNR nanogap device. We answered the question above through the $G_0$ spectra in **Figure 4** for all nucleobases and the whole rotation range (from 0° to 180°). It is evident from **Figure 4** that the sharp nucleobase's specific resonance peaks and the pattern remains throughout and are only being shifted to above/below the Fermi-level energies except for rS nucleobase for all the rotation angles. The picture is qualitatively the same for the rS nucleobase (**Figure 4(c)**). However, clearly distinct $G_0$ patterns and peaks are observed for B, P, S, and also to some extent for Z nucleobases (**Figure 4(a,b,d,e)**). Comparing these results for all the nucleobases leads to qualitative differences in the shift of the resonance peaks and the respective transmission peak energies as discussed above.

To disclose the impact of the rotation on the tunnelling current across the O-ZGNR nanogap+nucleobase, we present the individual results in **Figure 6**, from 0.1-1.0 V applied bias voltage and for all five nucleobases for all rotations. It is noted that these figure shows the nearly same quantities but different results as **Figure 6(a)**. All results presented in **Figure 6(a-g)** referred to simulations with a different rotation angle (i.e., 0° to 180° in steps of 30°) and applied bias voltage (0.1-1.0 V) for all the five nucleobases. At 0.4 V (refereeing to 120°), different current traces observed for atleast three nucleobases (i.e., P, S and Z) and can be arranged in the following sequence order: $I_Z > I_P > I_S$. When we increase the applied bias voltage from 0.4 to 0.8 V, the remaining two nucleobases (B and rS) could also be identified due to their different current traces. This suggest that we need atleast two different biases to identify all the five nucleobases while transported through the nanogap. Therefore, if experimental measurements pick up all the possible rotation angles, it may be able to distinguish these five nucleobases. Furthermore, it is interesting to note that in most of the rotation Z nucleobase transmits higher current compared to other four nucleobases. However, at 60°- and 90°- rotation, P nucleobase transmits a higher current among all the five nucleobases. This could be due to improved coupling and perfect tunnelling effect across the nanoelectrode+molecule. Also, these improvements of the electronic transverse current for certain rotations correlate to the $G_0$ resonance peaks shift, near to the Fermi-level under bias. These shifts may be larger for the nucleobases associated with the higher electronic tunnelling currents at a given rotation angle. Overall, the different rotation angles keep the distinct feature of the nucleobases in both electronic transverse $G_0$ spectra and tunnelling current features. In view of accurate DNA sequencing, both features could be utilized.

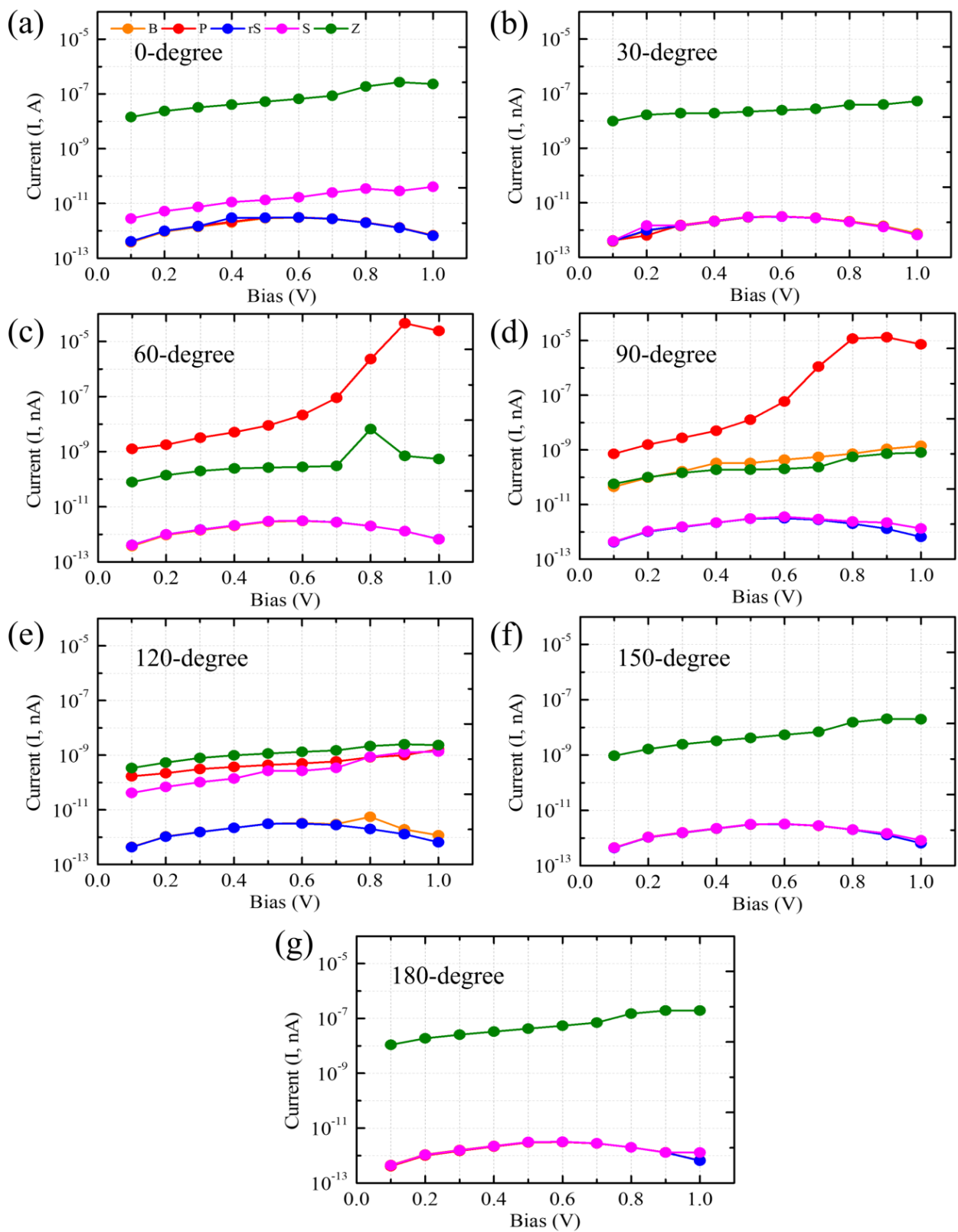

**Figure 6**. (a-g) The I-V characteristics of the nanogap+nucleobase (nucleobase: B, P, rS, S, and Z). The results are presented for all five nucleobases performing rotation within the nanogap in the yz-plane along the x-axis in the steps of 30° from 0° to 180° with respect to their original positions (0°).

**CONCLUSIONS**

We have discussed structural, electronic, and quantum transport features of the O-ZGNR device with and without Hachimoji nucleobases. The O-ZGNR device has shown quantum interference effects in the tunnelling conductance in the presence of at least four Hachimoji nucleobases. It has shown that the tunnelling conductance of Hachimoji DNA nucleobases at Fermi-level could be used to detect the nucleobase types. A variation of up to 2-5 orders of magnitude is observed in the conductance upon rotation. The maximum variation has found for the Z nucleobase conductance, which is significantly larger compared to the other four nucleobases. Moreover, it has been found that if we can control the rotation of the nucleobases, it may be possible to detect all the five nucleobases from their conductance values. Furthermore, our study indicates quantum interference effects such as anti-resonance or Fano-resonance that affect the variation of transverse conductance depending on the nucleobase conformation. The four nucleobases (B, P, S, and Z) shows that the LUMOs corresponding to transmission resonance peaks are localized on the *purine* and *pyrimidine* nucleobases. Therefore, the molecular and induced states contribute significantly to the conductance and I-V features. The I-V results suggested that all the five nucleobases may be possible to detect at two different applied bias voltages (i.e., 0.4 V and 0.8 V) if we can control the rotation angles of the nucleobases. Hence, if experimental measurements pick up all the possible rotation angles, it may be possible to distinguish among the five nucleobases upto 0.8 V bias. Thus, the different rotation angles keep the distinct feature of the nucleobases in both transverse conductance and tunnelling current features. We believe that developing such a nanoribbon-based device for sequencing of Hachimoji nucleobases should be feasible in principle within experimental realms.

**METHODS**

The modeled O-ZGNR nanogap device is shown in **Figure 1(b)** and **S1**. The nanogap with alternative OH- and H-terminated edges is computationally created with 1.26 nm size between the nanoelectrodes along the transport direction (z-axis). This has been done to ensure that the distance between target Hachimoji DNA nucleobase and the nanoelectrode is larger than the van der Waals (vdW) radius of hydrogen (H) and carbon (C) atoms (~3.0 Å). All five Hachimoji DNA nucleobases are placed between the nanoelectrodes such that the center of mass (COM) of each target nucleobases is located at approximately the middle of the O-ZGNR nanogap device.

The electronic structure computations of all the five Hachimoji DNA nucleobases are done using the B3LYP hybrid functional and 6-31+G* basis set level of theory with the Gaussian09 code.[31] The O-ZGNR nanoelectrode relaxation computations are carried out using the SIESTA code.[32,33] The vdW-DF-cx (van der Waals density functional consistent exchange) method, a double-zeta polarized (DZP) basis set, and norm-conserving Troullier-Martin pseudopotential is used in all simulations.[34–36] The vdW-DF-cx functional is well known to describe the weak interactions more accurately than the GGA-PBE calculations. A real-space cutoff of 200 Ry and a $1\times3\times2$ of *k*-point mesh in the reciprocal space based on the Monkhorst-Pack scheme are used in all electronic structure computations.

Further, each Hachimoji DNA nucleobase depicted in **Figure 1(a)** is separately positioned inside the nanogap of the O-ZGNR device so that the nucleobase and device planes are aligned. The nucleobases are placed in such a way that the nucleobase sites which take part in weak H-bonding in DNA strands are in the area of the membrane of the proposed device. After relaxation, each of the five nucleobases could assume the optimal electronic coupling configuration with respect to the graphene membrane. The fully optimized configurations (O-ZGNR+nucleobase) of all five setups are shown in **Figure S1(a-e)**.

To evaluate the energetic stability of the nucleobases in the O-ZGNR device, we have computed the $E_i$ values using the equation below:

$$E_i = [(E_{O-ZGNR} + E_{nucleobase}) - E_{O-ZGNR+nucleobase}] \qquad (1)$$

here $E_{O-ZGNR}$, $E_{nucleobase}$, and $E_{O-ZGNR+nucleobase}$ are the total energy of isolated O-ZGNR, and target nucleobase molecule, and the O-ZGNR with target nucleobase molecule inside the nanogap device, respectively.

The charge density difference (CDD) plots analysis is done to understand the charge-transfer-interaction between molecule and nanoelectrode edges by using the below-given equation:

$$\Delta\rho(r) = [\rho_{O-ZGNR+nucleobase}(r) - (\rho_{O-ZGNR}(r) + \rho_{nucleobase}(r))] \qquad (2)$$

here $\rho_{O-ZGNR}(r)$, $\rho_{nucleobase}(r)$, and $\rho_{O-ZGNR+nucleobase}(r)$ are the total charge density of isolated O-ZGNR, and target nucleobase molecule, and the O-ZGNR with target nucleobase molecule inside the nanogap device, respectively.

The quantum transport computations are achieved by combining the NEGF method with DFT using the TranSIESTA code.[33,33,37,38] The proposed O-ZGNR nanogap device is made up of two semi-infinite nanoelectrodes (left (L) and right (R)) and the scattering (device) region, as shown in **Figure 1(b)** and **S1**. The quantum transport computations are performed along the z-direction, and the L/R nanoelectrode plays the role of the source/drain of nanoelectrodes, respectively. The basis set used in the transport computation is the same as those used in the electronic structure computation. The $1 \times 11 \times 1$ *k*-points are used to define the real-space grid. Using the DFT+NEGF approach, we compute the transmission function $[T(E)]$ at zero-bias and current characteristics. The $T(E)$ is computed using the following equation:

$$T(E) = tr[\Gamma_L(E)G_C(E)\Gamma_R(E)G_C^\dagger(E)] \qquad (3)$$

where $\Gamma_{L/R}(E)$ represents coupling matrix of $L$ and $R$ nanoelectrodes, and $G_C(E)/G_C^\dagger(E)$ are the retarded and advanced Green's functions. At zero-bias, we have computed the conductance $G = G_0\, T(E_F)$, where $G_0 = 2e^2/h$ represents the quantum conductance, $e$ (charge of the electron) and $h$ (Plank's constant); respectively.[3,4,8,8,9,29,37–40]

The integration of equation 3 provides the current i.e., $I(V_b)$ (current under applied bias voltage ($V_b$)), which can be computed by using the below equation:

$$I(V_b) = \frac{2e}{h} \int_{\mu_R}^{\mu_L} T(E)\,[f(E - \mu_L) - f(E - \mu_R)]\,dE \qquad (4)$$

where $f(E - \mu_L)$ and $f(E - \mu_R)$ represents the Fermi-Dirac functions for the electrons in the $L$ and $R$ nanoelectrodes, respectively.[37,38,41–45]

## ASSOCIATED CONTENT

**Supporting Information:** Geometries; rotation schemes; and electronic density of states plots.

## AUTHOR INFORMATION

### Corresponding Author

Rameshwar L. Kumawat − Department of Chemistry, Indian Institute of Technology (IIT) Indore, Indore, Madhya Pradesh 453552, India; orcid.org/0000-0002-2210-3428; Email: rameshwarlal1122@gmail.com

Biswarup Pathak − Department of Chemistry, Indian Institute of Technology (IIT) Indore, Indore, Madhya Pradesh 453552, India; orcid.org/0000-0002-9972- 9947; Email: biswarup@iiti.ac.in

Author

## ACKNOWLEDGMENTS

We acknowledge IIT Indore for the lab and computing facilities. This work is supported by SPARC/2018-2019/P116/SL and DST-SERB (Project Number: CRG/2018/001131), and CSIR (project 01(3046)/21/EMR-II).

## CONFLICTS OF INTEREST

No conflicts of interest to declare.

**Table of Content Graphics:**

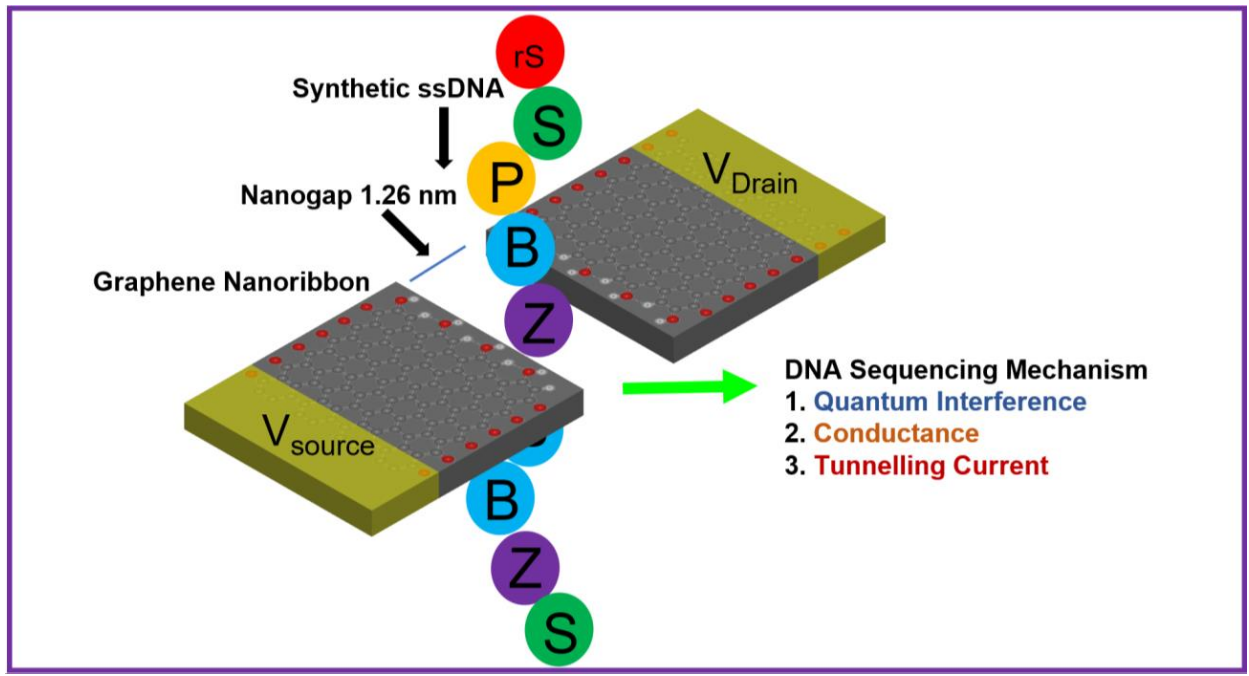